\documentclass[fleqn,usenatbib]{mnras}
\usepackage{graphicx,pdflscape}

\def\EBV{\mbox{$E(B-V)$}}
\def\dBV{\mbox{$\delta E(B-V)$}}

\def\Zo{\mbox{$Z/Z_\odot$}}

\def\ms{\mbox{$M_\odot$}}
\def\ds{\mbox{$d_\odot$}}

\def\x2{\mbox{$R_H$}}
\def\fH{\mbox{$[Fe/H]$}}
\def\fC{\mbox{$fitCMD$}}
\def\fC{\textit{fitCMD}}

\def\rT{\mbox{r$_{T}$}}
\def\rC{\mbox{r$_{C}$}}
\def\mB{\mbox{$m_{\rm F606W}$}}
\def\mR{\mbox{$m_{\rm F814W}$}}

\title[Seeing through the dust of Palomar\,2]{Lifting the dust veil from the Globular Cluster 
Palomar\,2}

\author[C. Bonatto and A.L. Chies-Santos]{Charles Bonatto and Ana L. Chies-Santos\\
Departamento de Astronomia, Universidade Federal do Rio Grande do Sul, Av. Bento
Gon\c{c}alves 9500\\
Porto Alegre 91501-970, RS, Brazil\\ }

\begin{document}

\maketitle

\begin{abstract}

This work employs high-quality {\em Hubble Space Telescope} ({\em HST}) Advanced Camera for Surveys (ACS) 
F606W and F814W photometry to correct for the differential reddening affecting the colour-magnitude 
diagram (CMD) of the poorly-studied globular cluster (GC) Palomar\,2. Differential reddening is taken 
into account by assuming that morphological differences among CMDs extracted across the field of view 
of Palomar\,2 correspond essentially to shifts (quantified in terms of \dBV) along the reddening vector 
due to a non-uniform dust distribution. The average reddening difference over all partial CMDs is 
$\overline{\dBV}=0.24\pm0.08$, with the highest reaching $\dBV=0.52$. The corrected CMD displays
well-defined and relatively narrow evolutionary sequences, especially for the evolved stars, i.e. the 
red-giant, horizontal and asymptotic giant branches (RGB, HB and AGB, respectively). The average width 
of the upper main sequence and RGB profiles of the corrected CMD corresponds to 56\% of the original 
one. Parameters measured on this CMD show that Palomar\,2 is $\approx13.25$\,Gyr old, has the mass 
$M\sim1.4\times10^5\,\ms$ stored in stars, is affected by the foreground $\EBV\approx0.93$, is located 
at $\ds\approx26$\,Kpc from the Sun, and is characterized by the global metallicity $Z/Zo\approx0.03$, 
which corresponds to the range $-1.9\leq\fH\leq-1.6$ (for $0.0\leq[\alpha/Fe]\leq+0.4$), quite consistent 
with other outer halo GCs. Additional parameters are the absolute magnitude $M_V\approx-7.8$, and the 
core and half-light radii $\rC\approx2.6$\,pc and $R_{HL}\approx4.7$\,pc, respectively.

\end{abstract}

\begin{keywords}
{{\em (Galaxy:)} globular clusters: general; ({\em Galaxy}:) globular clusters: individual: Palomar\,2} 
\end{keywords}

\section{Introduction}
\label{intro}

Structure formation in the Universe is described as hierarchical by the $\Lambda-$cold dark matter 
($\Lambda-$CDM) cosmology, with smaller pieces merging together to build up the larger galaxies 
currently observed. In the Milky Way, several instances of accretion have been recently detected, 
such as the cases of the Sagittarius dwarf spheroidal galaxy (\citealt{Ibata94}), the halo stellar 
streams crossing the solar neighbourhood (\citealt{Helmi99}), and the stellar debris from Gaia-Enceladus 
(\citealt{Helmi2018}, \citealt{Belokurov2018}). In fact, as new deep and wide-area photometric surveys 
are being conducted, new stellar streams are being found. For instance, 11 new streams have been recently 
discovered in the Southern sky (\citealt{Shipp2018}) by the Dark Energy Survey (DES - \citealt{Bechtol2015}). 
In summary, more than 50 stellar streams have been found relatively recently, half of which were discovered 
in the short period after 2015 (e.g. \citealt{Mateu2018}; \citealt{Li2019}). 

Accretions and mergers (mainly of dwarf galaxies) are expected to have added not only field stars, 
but also entire Globular clusters (GCs) to the Milky Way stellar population (\citealt{Penarrubia2009}). 
Indeed, based on differences present in the age $\times$ metallicity relation of a large sample of Milky 
Way GCs, \citet{ForbesBridges2010} estimate that $\approx25\%$ of them have been accreted from perhaps 
6-8 dwarf galaxies. On the other hand, also focusing on the age $\times$ metallicity relation,
\citet{Leaman2013} suggests that accretion is responsible for all of the halo Milky Way GCs. More
recently, working with Gaia (\citealt{Gaia}) kinematic data for 151 Galactic GCs, \citet{Massari2019} 
found that $\approx40\%$ of the present-day clusters likely formed in situ, while $\approx35\%$ appear 
to be associated with known merger events. These works show that the origin of the Galactic GC system
has not yet been settled, and point to the fact that in-depth analysis of each one of the GCs is
important.

Currently, about 160 Milky Way GCs are known (e.g. \citealt{Harris2010}), and most of them have 
already been the subject of detailed photometric, spectroscopic and/or kinematical studies with the 
multiple goals of determining their collective properties as a family, and to find their individual 
origin (formed in situ or accreted), among others. This census is expected to increase - especially 
towards the faint end of the GC distribution, as deep photometric surveys become available. For 
instance, working with photometry from the VISTA Infrared Camera (VIRCAM - \citealt{Emerson2010}) at 
the 4m VISTA telescope at the ESO Cerro Paranal Observatory, \citet{Palma2019} present 17 new candidates 
to Bulge GCs. Having being formed - or accreted - at the early 
stages of the Galaxy, determining the present-day parameters of all the individual GCs, together with 
their large-scale spatial distribution, may provide important clues to the Milky Way assembly process 
as well as to the chemical and physical conditions then prevailing. In addition, individual GCs have 
been used as test-beds for dynamical and stellar evolution theories (e.g. \citealt{Li2016RAA}; 
\citealt{Chatterjee2013}; \citealt{Kalirai2010}). 

Despite having been discovered in the 1950's in survey plates from the first National Geographic 
Society – Palomar Observatory Sky Survey, Palomar\,2 remains one of the least studied Milky Way 
GCs. The main reason for this is that, although located far from the Galactic bulge at $\ell=170.53\degr$ 
and $b=-9.07\degr$, there is a foreground thick wall of absorption from the Galactic disk that makes 
it a relatively faint cluster, with $V\sim13$, and leads to severe photometric scattering related to
differential reddening. This feature is clearly present in the earliest $V\times(V-I)$ colour-magnitude 
diagram (CMD) of Palomar\,2 obtained with the UH8K camera at the Canada-France-Hawaii Telescope (CFHT) 
by \citet{Harris97}. Besides the differential reddening, they describe the CMD of Palomar\,2 as showing 
a well-populated red horizontal branch with a sparser extension to the blue, similar to the CMDs of the 
GCs NGC\,1261, NGC\,1851, or NGC\,6229. Nevertheless, by comparison with CMDs of other GCs, they were 
able to estimate the reddening $\EBV=1.24\pm0.07$, the intrinsic distance modulus of $(m-M)_0=17.1\pm0.3$ 
thus implying a distance from the Sun of $\ds\approx27$\,Kpc, and a metallicity $\fH\approx-1.3$, 
characterizing it as an outer halo GC. They also estimated its integrated luminosity as $M_V\approx-7.9$, 
thus making it brighter - and consequently more massive - than most other clusters in the outer halo. 

More recently, Palomar\,2 was included in the photometric sample obtained under program number GO\,10775 
(PI: A. Sarajedini) with the {\em Hubble Space Telescope} ({\em HST}) Advanced Camera for Surveys (ACS). 
GO\,10775 is a {\em HST} Treasury project in which 66 GCs were observed through the F606W ($\sim V$) and 
F814W ($\sim I$) filters. Data reduction and calibration into the VEGAMAG system were done by
\citet{Anderson2008}. However, because of the differential-reddening related photometric scatter,  
Palomar\,2 was essentially excluded from subsequent CMD analyses.

In this paper, the calibrated photometric catalog containing F606W and F814W magnitudes are used 
to correct the CMD of Palomar\,2 for the deleterious effects of differential reddening, leading to a 
robust determination of its intrinsic parameters.

This paper is organised as follows: Sect.~\ref{TOM} describes the differential-reddening analysis 
and presents the corrected CMD; intrinsic parameters of Palomar\,2, such as the total stellar mass,
age, metallicity, and distance to the Sun, will be obtained in Sect.~\ref{fPar}. Concluding remarks 
are given in Sect.~\ref{CONC}.

\section{Differential reddening towards Palomar\,2}
\label{TOM}

A previous analysis of the differential reddening in Palomar\,2 was undertaken by \citet{Sarajedini2007} 
with the F606W and F814W photometry obtained from the {\em HST} WFC/ACS Treasury project GO\,10775; observations 
and data reduction are fully described in \citet{Anderson2008}. Considering the difficulties caused by
the differential reddening, they simply extracted radial CMDs in 4 rings around the GC center containing 
stars affected by different reddening values. This allowed them to produce a somewhat low-spatial resolution 
correction to the original photometry (their Figs.~19 and 20). Then, by comparing the latter CMD with the 
fiducial line of the GC NGC\,6752, they estimated a metallicity in the range $-1.68\leq\fH\leq-1.42$, the
foreground reddening $\EBV=0.94$, and the intrinsic distance modulus $(m-M)_0=17.13$, and assumed that the 
age of Palomar\,2 should be similar to that of NGC\,6752.

In the present work, differential reddening in Palomar\,2 is dealt with by means of properties 
encapsulated in its $(\mB-\mR) \times\mB$ CMD. 
Colour-magnitude diagrams are one of the best tools with which to extract intrinsic parameters (e.g. 
age, metallicity, mass distribution and the shape of the Horizontal Branch) of a stellar population. 
This task can be accomplished by interpreting the morphology of evolutionary sequences on CMDs with the 
latest and more complete sets of theoretical isochrones. However, this kind of analysis depends heavily
on the photometric quality and consistency of the CMDs. Scatter, either extrinsic (e.g. related to low 
exposure time and/or ground-based telescopes) or intrinsic (coming from a non-uniform distribution of 
dust and/or large amounts of field-star contamination), may turn a CMD virtually useless for a rigorous 
analysis. 

\begin{figure}
\resizebox{\hsize}{!}{\includegraphics[width=\textwidth]{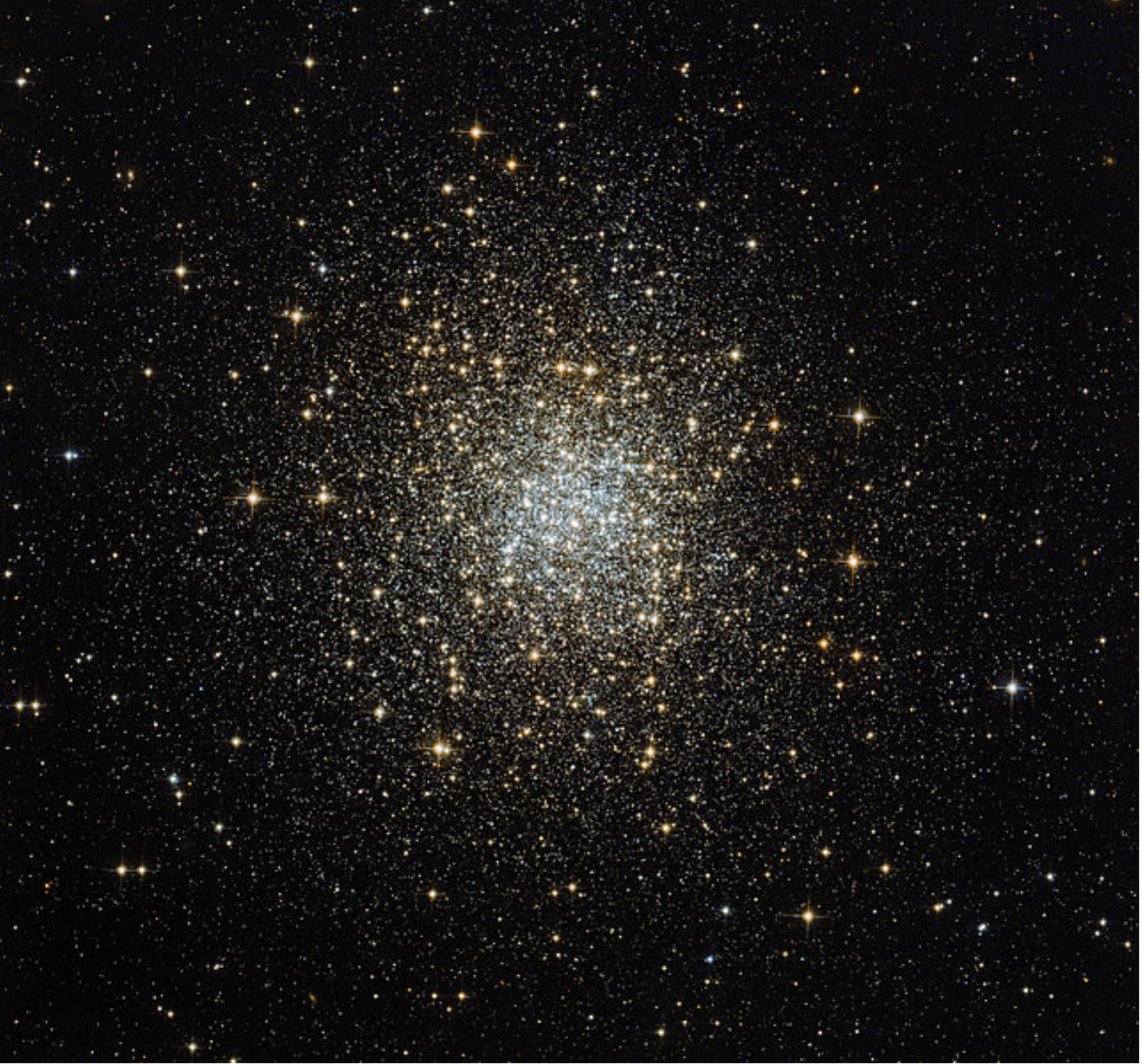}}
\caption[]{Palomar\,2 F606W ($\sim V$) and F814W ($\sim I$) composite image by ESA/Hubble \& NASA. 
North is up and East to the left.}
\label{fig1}
\end{figure}

According to \citet{Harris2010}, the half-light and tidal radii of Pal\,2 are $\approx30\arcsec$
and $\approx400\arcsec$, respectively. Thus, although far from reaching the tidal radius, the 
$\approx200\arcsec\times200\arcsec$ field of view of WFC/ACS surely samples most of the member 
stars, as can be seen in Fig.~\ref{fig1}. 

The approach to map - and correct for - the differential reddening basically follows the method 
described in \citet{DifRed12}. In summary, in that work the field of view (FoV) of a star cluster 
was divided in a grid of (fixed-size) cells, from which partial CMDs were extracted and compared 
with the average CMD (i.e. with all stars in the field). Then it was assumed that morphological 
differences among the partial CMDs were essentially linear shifts along the reddening vector 
due to non-uniform dust distribution. In this context, the approach computed the value of \EBV\ 
that made each partial CMD agree with the average one. Finally, a map was produced for \dBV, 
the difference in \EBV\ between each partial CMD and the bluest (that with the least amount of
reddening) one. 

Although the use of fixed-size cells simplifies the algorithm, in general it leads to oversampling 
the inner regions, thus lowering the spatial resolution there. Since the stellar surface density
of GCs roughly follows a power-law with a core (e.g. \citealt{King62}), same-size cells located 
near the GC center are bound to contain many more stars than those at the outskirts. So, if a 
minimum number of stars is required to define a CMD, obviously this criterion would be met by a
much smaller cell near the center than outwards. Consequently, an improvement to the original approach 
was introduced so that now variable-size cells are used. 

\begin{figure}
\resizebox{\hsize}{!}{\includegraphics[width=\textwidth]{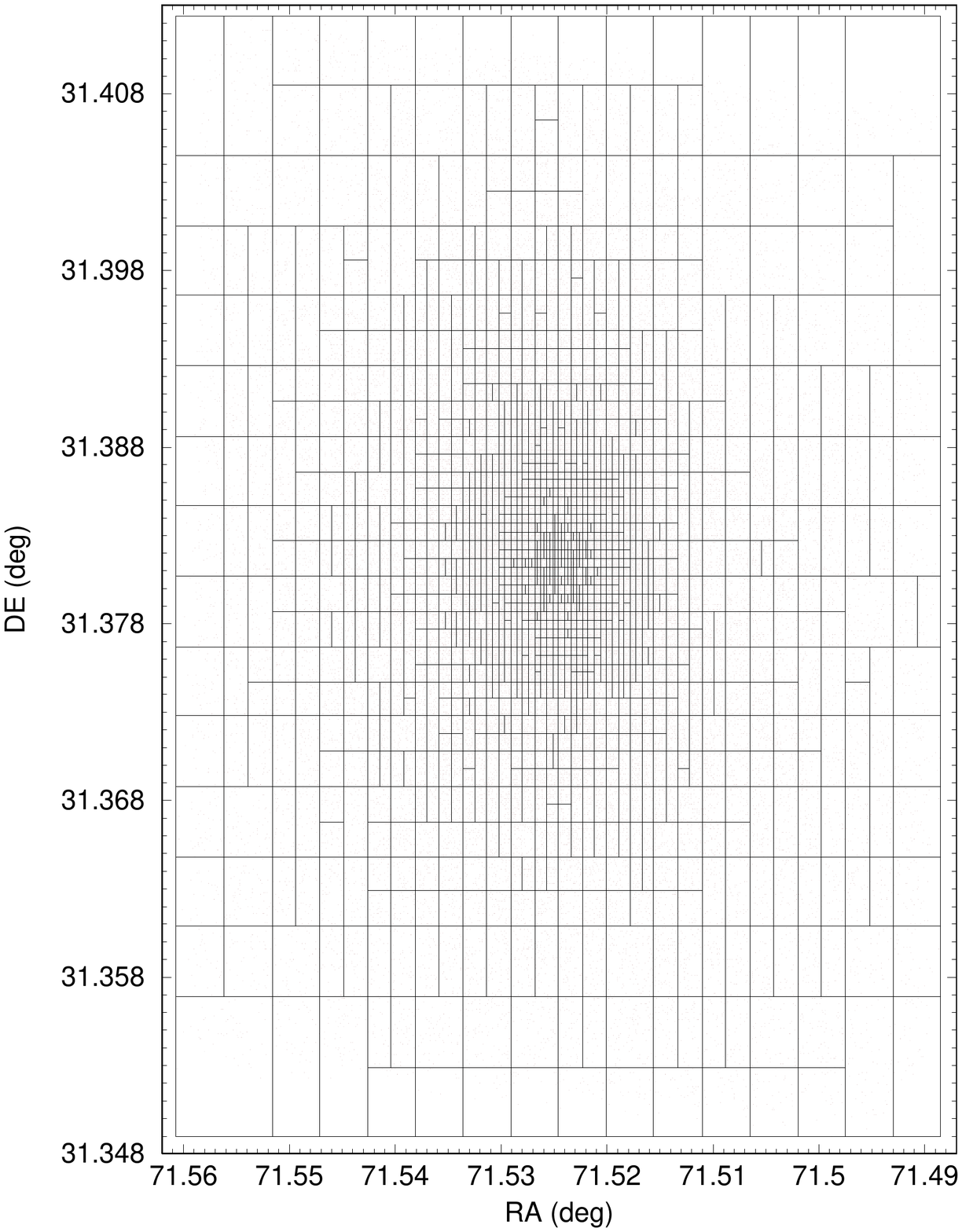}}
\caption[]{Distribution of the partial CMDs (black polygons) overplotted on the {\em HST} WFC/ACS
photometry (brown points). The number-density of extracted CMDs increases towards the center 
following the GC stellar surface density.}
\label{fig2}
\end{figure}

First one sets the minimum number ($N_{min}$) of stars required to roughly define a CMD (this value 
depends on the quality of the photometry), and the field of view of the GC is divided into 4 quadrants
to be individually analysed. Then, if the number of stars in any quadrant is higher than $4\times N_{min}$, 
a new division by 4 is applied to that quadrant, and so on recursively, until the number of stars in the 
higher order quadrants gets lower than $4\times N_{min}$. The partial CMDs are then built only for the 
quadrants having at least $N_{min}$ stars. Figure~\ref{fig2} shows that the number-density of cells  
basically follows the spatial density of stars across the field of Palomar\,2, which increases towards 
the central parts. Also, the cell dimension decreases towards the central parts because all cells are
required to include a minimum number of stars.

Instead of the discrete CMDs, here we use Hess (\citealt{Hess24}) diagrams, their continuous 
analogues that take photometric uncertainties into account. The values of \dBV\ are found by
minimizing the residuals of the difference between the average and partial Hess diagrams.
Minimization is achieved by means of the global optimisation method Simulated Annealing (SA, 
\citealt{SA94}; \citealt{BoLiBi2012}) is used to compute \dBV\footnote{SA originates from the 
metallurgical process by which the controlled heating and cooling of a material is used to increase 
the size of its crystals and reduce their defects. If an atom is stuck to a local minimum of the 
internal energy, heating forces it to randomly wander through higher energy states.}.

\begin{figure}
\resizebox{\hsize}{!}{\includegraphics[width=\textwidth]{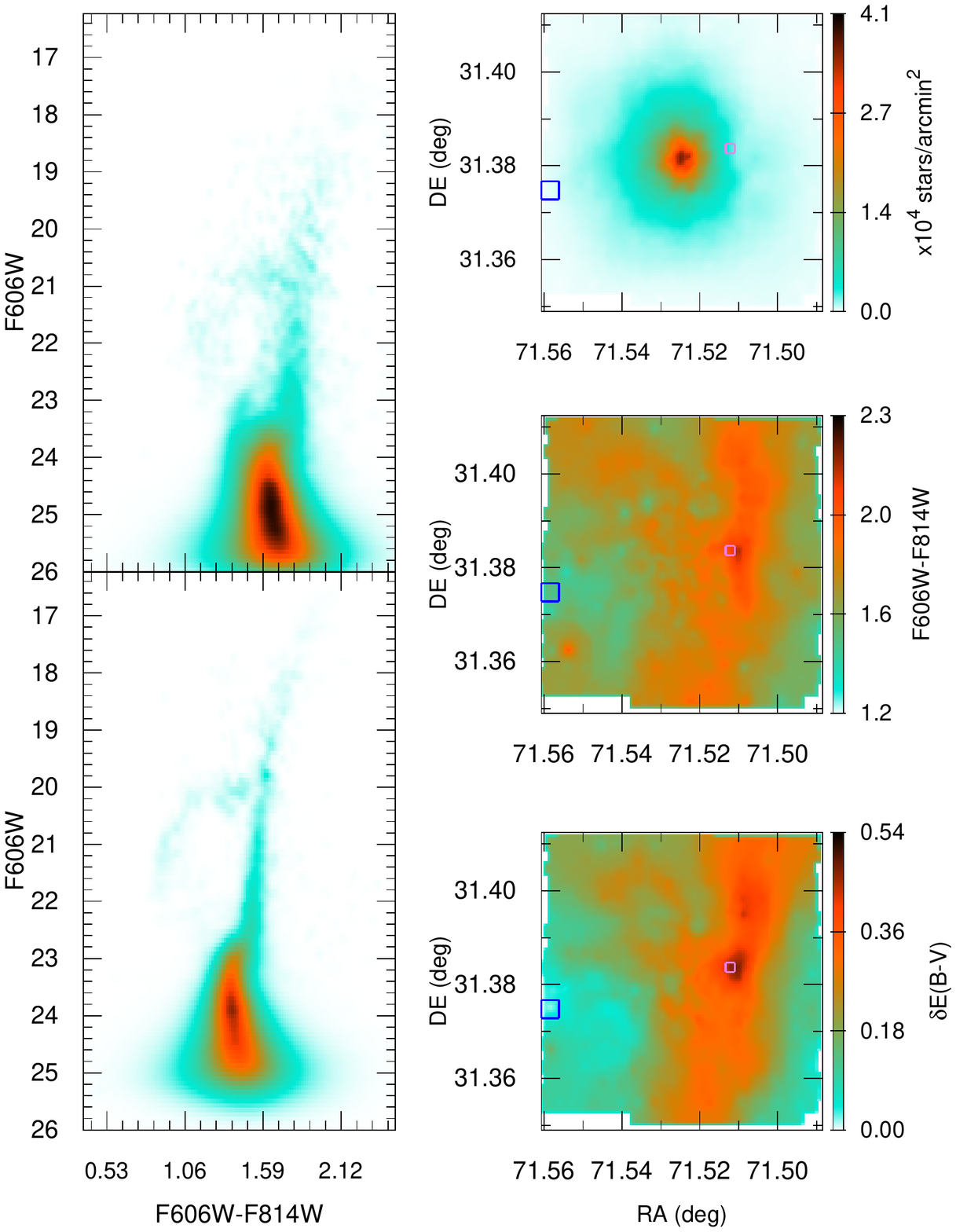}}
\caption[]{Original (top-left panel) and differentially-reddening corrected photometry (bottom-left)
of Palomar\,2. Right panels show the stellar surface density (top), colour (middle) and \dBV\ (bottom)
spatial distributions. The blue and violet rectangles locate the bluest and reddest cells, respectively.}
\label{fig3}
\end{figure}

Once the \dBV\ map has been built, it is straightforward to correct the values of F606W and F814W
for each star and produce the differential-reddening corrected CMD (or Hess diagram). This
process is illustrated in Fig.~\ref{fig3}. Clearly, the relatively heavy differential reddening in 
Palomar\,2 leads to a significant scatter and broadening of the upper main sequence and the evolved 
sequences - red-giant branch (RGB), horizontal branch (HB) and asymptotic giant branch (AGB) - (Top-right 
panel). After correction, the scatter gets much reduced and the sequences become better defined 
(bottom-left). Right panels show the spatial distribution of the density of stars (top), the colour 
$\mB-\mR$ (middle) and \dBV\ (bottom). 

\begin{figure}
\resizebox{\hsize}{!}{\includegraphics[width=\textwidth]{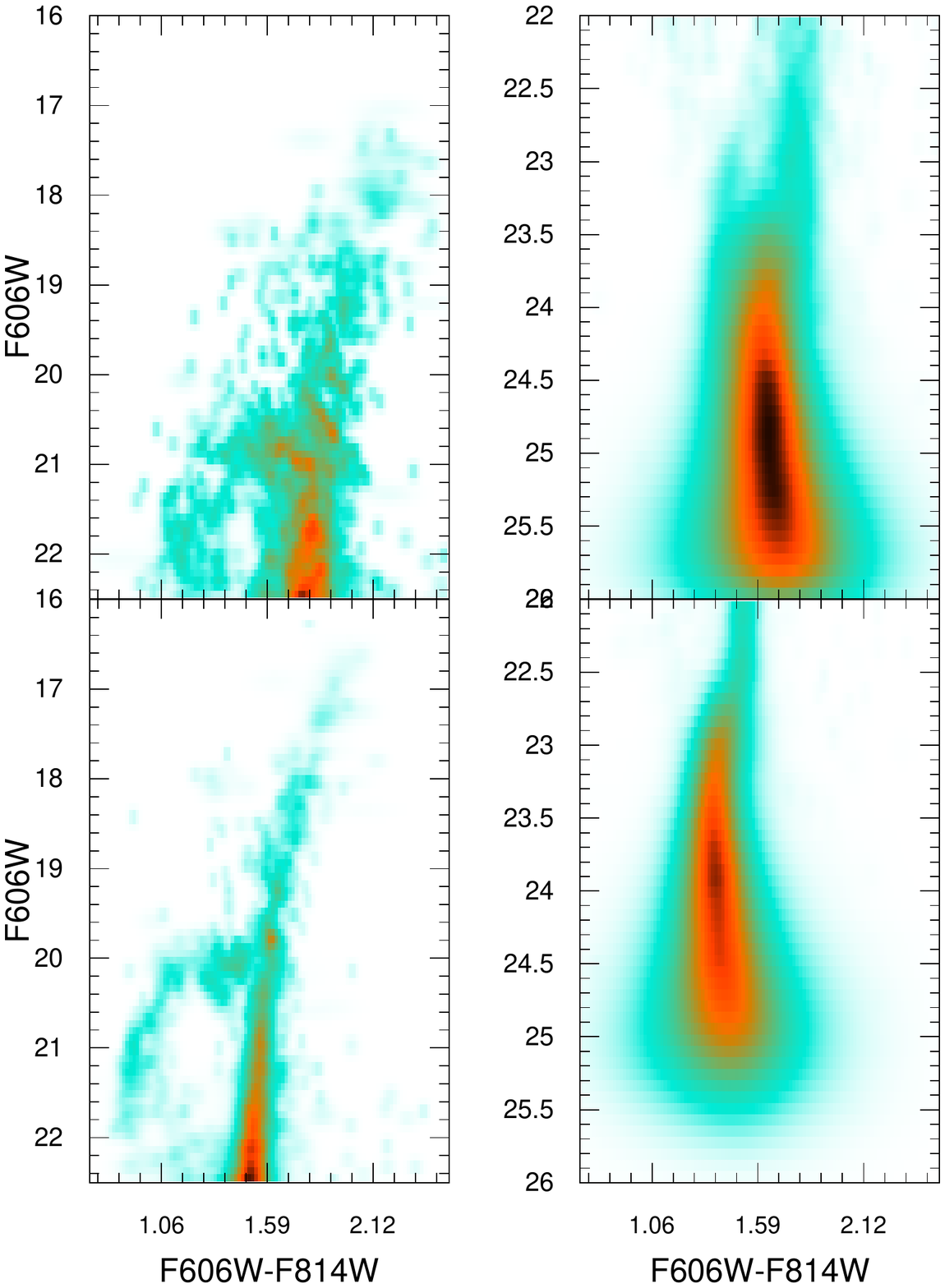}}
\caption[]{Left panels compare the original (top) and differentially-reddening corrected evolved
stellar sequences RGB, HB and AGB (bottom); Right: same for the upper main sequence.}
\label{fig4}
\end{figure}

As expected, the spatial distribution of \dBV\ essentially matches that of the colour, especially 
the nearly South-North dust lane. The location and size of the bluest and reddest CMDs are also shown. 
The reddest CMD ($\dBV=0.52$) is located at $RA = 71.5122\degr = 04^{\rm h}$:$46^{\rm m}$:$2.9^{\rm s}$ 
and $DE = 31.3839\degr = +31\degr$:$23^{\rm m}$:$2^{\rm s}$, while the bluest is at $RA = 71.5583\degr = 
04^{\rm h}$:$46^{\rm m}$:$14^{\rm s}$ and $DE = 31.3750\degr = +31\degr$:$22^{\rm m}$:$30^{\rm s}$, near 
the border of the WFC/ACS field. The average value of the differential reddening across the WFC/ACS field 
is $\overline{\dBV}=0.24\pm0.08$. Such a high value is consistent with the intricate and broad evolutionary 
sequences seen in the original CMD of Palomar\,2 (Fig.~\ref{fig3}).



Changes imparted on the stellar sequences by the differential-reddening correction can be better 
appreciated by zooming in on specific CMD ranges. As shown in Fig.~\ref{fig4}, the evolved sequences
(left panels) previously barely discernible, now become well defined and tight, displaying beautiful
RGB, AGB and a blue-HB, typical of metal-poor GCs. Improvements to the upper main sequence are also
quite apparent, especially a better definition of the turn-off (right panels). To better appreciate 
this improvement, we build the average profiles along the upper main sequence and RGB before and 
after differential-reddening correction showing the profile amplitude (normalized to the peak value) 
as a function of the difference in colour ($\delta_{\rm colour}$) with respect to the mean-ridge line at 
a given magnitude (see Sect.~\ref{fPar} for further details). The average profile width measured across   corresponds to 56\% of the original profile. 

\section{Fundamental parameters of Palomar\,2}
\label{fPar}

The differential-reddening corrected photometry in the previous Section can now be used to obtain robust 
fundamental parameters by means of the approach \fC, fully described in \citet{fitCMD}. In summary, the 
rationale underlying \fC\ is to transpose theoretical initial mass function (IMF) properties - encapsulated 
by isochrones of given age and metallicity - to their observational counterpart, the CMD. This requires 
finding values of the total (or cluster) stellar mass, age, global metallicity (Z), foreground reddening, 
apparent distance modulus, as well as for parameters describing magnitude-dependent photometric completeness. 
These parameters - including the actual photometric scatter taken from the photometry - are used to build 
a synthetic CMD that is compared with that of a star cluster. Residual minimization between observed and 
synthetic CMDs - by means of the global optimization algorithm Simulated Annealing - then leads to the 
best-fit parameters.

\begin{figure}
\resizebox{\hsize}{!}{\includegraphics[width=\textwidth]{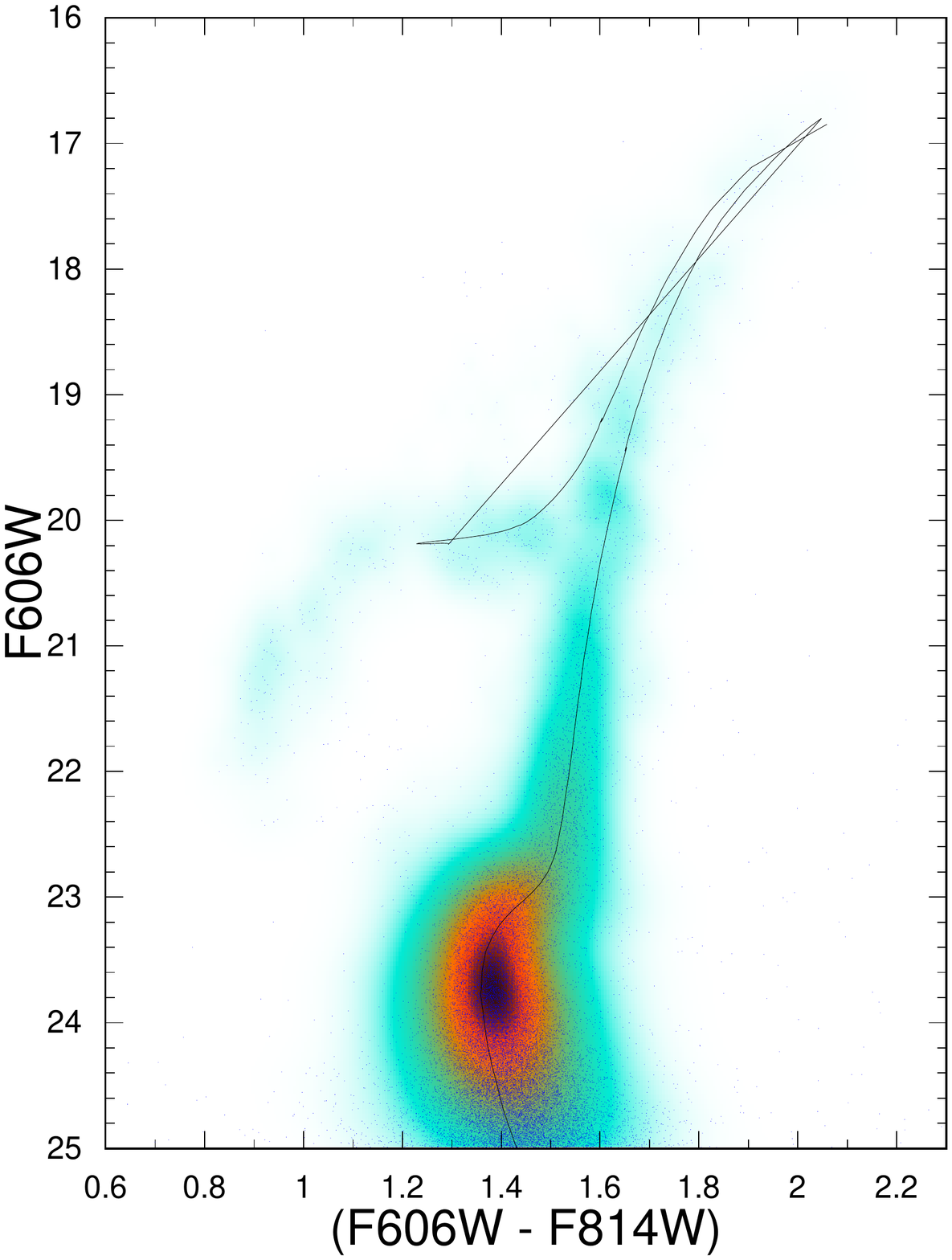}}
\caption[]{Isochrone fit to the differential-reddening corrected CMD of Palomar\,2 according to \fC. 
The best-fit isochrone parameters are: $13.25$\,Gyr of age and total metallicity $Z=4\times10^{-4}$. 
The actual stars are superimposed on the Hess diagram.}
\label{fig5}
\end{figure}

\begin{figure}
\resizebox{\hsize}{!}{\includegraphics[width=\textwidth]{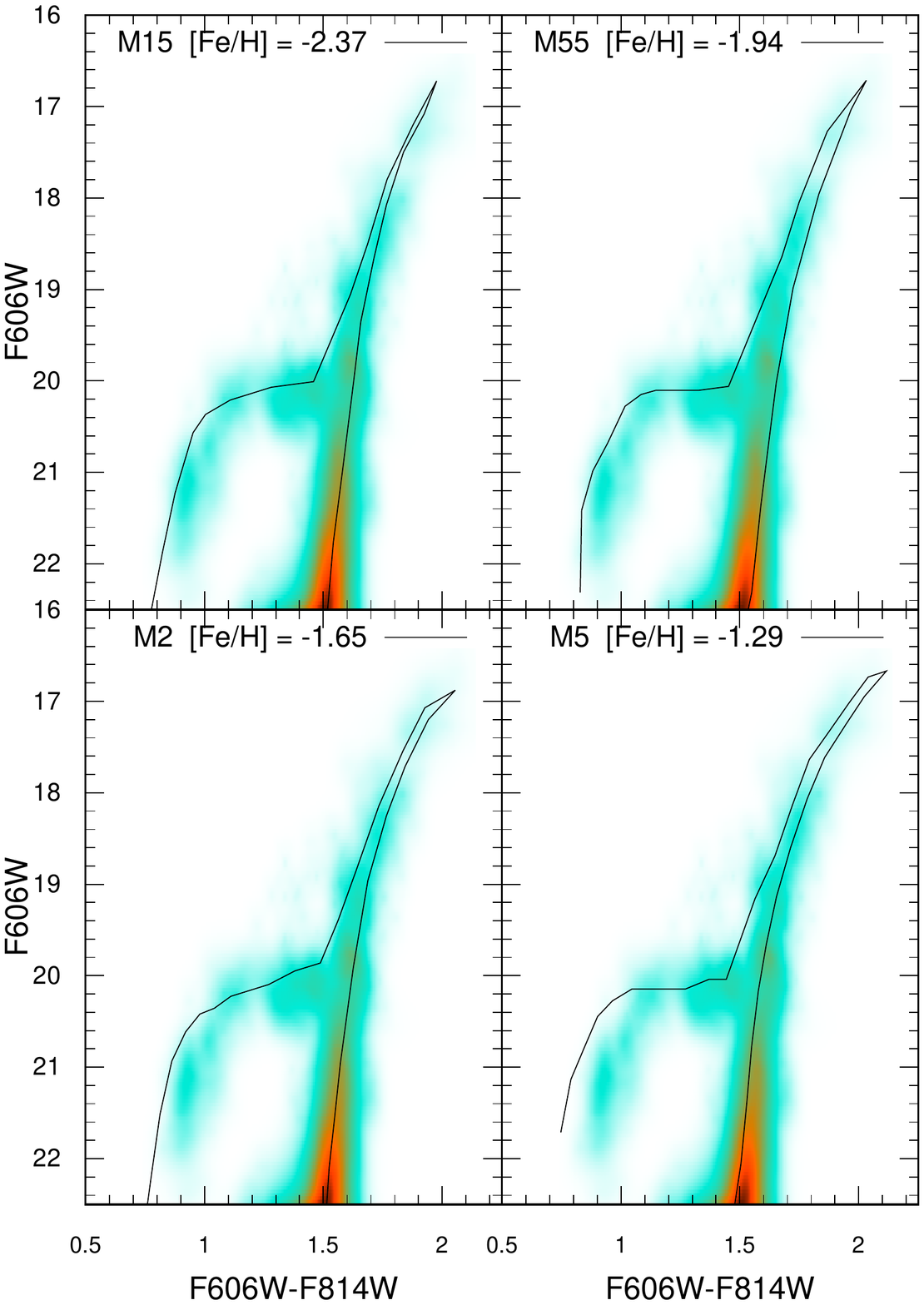}}
\caption[]{Comparison of the differential-reddening corrected evolved sequences of Palomar\,2
with the mean-ridge lines of GCs with different metallicity.}
\label{fig6}
\end{figure}

Regarding isochrones, \fC\ employs the latest PARSEC v1.2S$+$COLIBRI PR16 (\citealt{Bressan2012};
\citealt{Marigo2017}) models\footnote{Downloadable from $http://stev.oapd.inaf.it/cgi-bin/cmd$}.
These isochrones deliver a relatively comprehensive set of fundamental physical parameters for 
each stellar mass (from the Hydrogen-burning limit to the mass corresponding to stars in highly 
evolved stages, but still observable in CMDs). Thus, the PARSEC models are quite adequate to
build artificial CMDs representative of actual stellar populations.

Since the differential-reddening corrected Hess diagram presents well-defined evolutionary sequences 
and relatively low photometric scatter, \fC\ readily converged to the best-fit parameters total stellar 
mass $M=(14\pm4)\times10^4\,\ms$, the foreground reddening $\EBV=0.94\pm0.03$, the intrinsic distance 
modulus $(m-M)_0=17.08\pm0.13$ (within the uncertainties, both values agree with those given by \citet{Sarajedini2007}, the distance to the Sun $\ds=26.1\pm1.5$\,Kpc, and the absolute magnitude 
$M_{\rm F606W}=-7.8$ (which essentially corresponds to $M_V$). Within the uncertainties, the value of \ds\ 
agrees with the $27.2$ distance found by \citet{Harris2010}; the same applies to $M_V$. The global 
metallicity is $Z=(4\pm1)\times10^{-4}$, which corresponds to $\Zo\approx0.03$. Since the specific 
value of $[\alpha/Fe]$ is not known for Palomar\,2, we consider the range $0.0\leq[\alpha/Fe]\leq+0.4$
which is typical for outer halo GCs (e.g. \citealt{DSA11}). Thus, the metallicity of Palomar\,2
should be in the range $-1.91\leq\fH\leq-1.58$ (with an uncertainty of $\pm0.08$. This value is 
considerably more metal poor than previously quoted by \citet{Harris2010}, but fully consistent
with the values found for other outer halo GCs (\citealt{DSA11}). The best-fit age is 
$13.25\pm0.12$\,Gyr, again consistent with other outer halo GCs.  


The best-fit isochrone (13.25\,Gyr of age and $Z=4\times10^{-4}$) set with the apparent distance modulus
$(m-M)_{\rm F606W}=19.73$ and the foreground reddening $E(\rm F606W-F814W)=0.93$ is shown in Fig.~\ref{fig5}, 
superimposed both on the Hess diagram and CMD of Palomar\,2. Although PARSEC isochrones lack the blue-HB 
extension, they still provide an adequate fit to the upper main sequence, RGB and AGB. 

\begin{figure}
\resizebox{\hsize}{!}{\includegraphics[width=\textwidth]{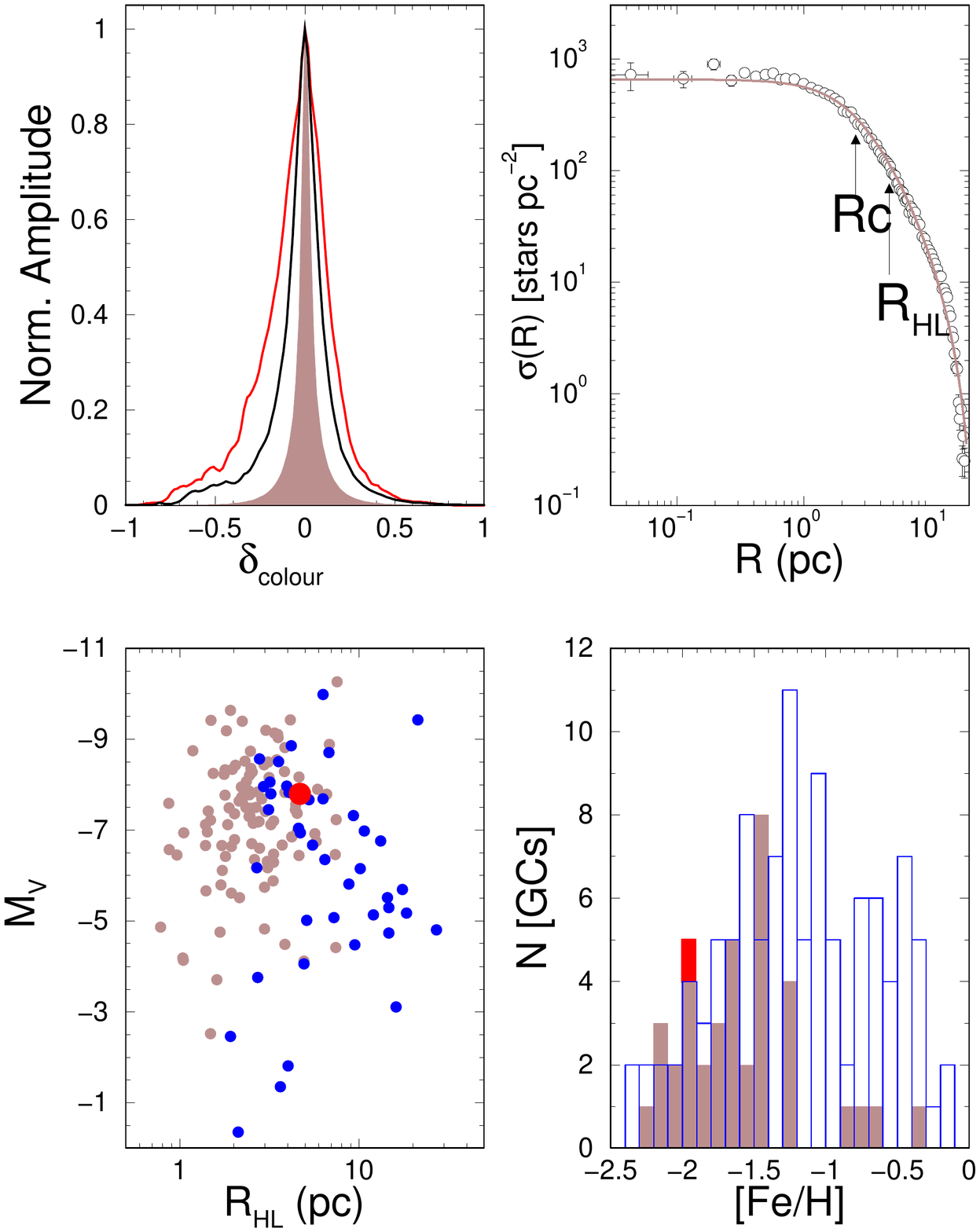}}
\caption[]{Top-left: average profile along the upper-main sequence and RGB before (red line)
and after (black) differential-reddening correction; the profile corresponding to the 
photometric scattering is shown in brown. Left: radial density profile of Palomar\,2 
built with the ACS photometry fitted with the 3-parameter King profile (solid line); core and 
half-light radii are indicated. Bottom-left: $M_V \times R_{HL}$ for the outer-halo GCs (blue
symbols), the remaining GCs (brown), and Palomar\,2 (red). Right: metallicity of Palomar\,2 (red)
compared to the Milky Way GCs in the outer halo (blue) and the remaining GCs (brown). Since the actual 
value of $[\alpha/Fe]$ is not known for Palomar\,2, we show the range of possible \fH\ values 
corresponding to $0.0\leq[\alpha/Fe]\leq+0.4$.}
\label{fig7}
\end{figure}

An alternative way to assess the metallicity of Palomar\,2 is by comparing its differential-reddening 
corrected CMD with those of GCs covering a range in \fH. As quality criteria, the comparison GCs 
are taken from the same photometric sample as Palomar\,2, and their CMDs should have a relatively 
large number of stars and low reddening (to better define the stellar sequences). Objects satisfying 
these criteria are M\,15 (NGC\,7078), $\fH=-2.37$, $\EBV=0.1$; M\,55 (NGC\,6809), $\fH=-1.94$, 
$\EBV=0.08$; M\,2 (NGC\,7089), $\fH=-1.65$, $\EBV=0.06$; and M\,5 (NGC\,5904), $\fH=-1.29$, 
$\EBV=0.03$. In addition, to minimize clutter, the analysis is carried out with the mean-ridge 
lines of the evolved stellar sequences of the comparison GCs fitted to Palomar\,2.

The result is shown in Fig.~\ref{fig6}, where the metal-poor nature of Palomar\,2 is reinforced.
Indeed, the mean-ridge lines of both M\,15 and especially M\,55 provide an excellent description
of the evolved sequences of Palomar\,2, thus setting its metallicity to a value closer to
$\fH=-1.9$ than $\fH=-1.6$.

Incidentally, Fig.~\ref{fig7} (top-left panel) shows that the differential-reddening corrected 
profile, although significantly narrower ($\approx56\%$) than the observed one, is broader than that
corresponding to the photometric uncertainties alone. Such excess width might suggest the presence of 
multiple stellar populations in Palomar\,2 (for recent reviews on multiple stellar populations, see 
\citealt{Gratton2012} and \citealt{Bastian2017}). However, it should be noted that the photometric 
catalogs of \citet{Anderson2008} are affected by unaccounted-for telescope breathing effects, i.e. 
the change of the PSF shape from one exposure to the next, even within the same HST orbit. This 
change in the PSF shape can be as large as a few percent, thus introducing systematic photometric 
errors of about the same amount. In this context, further speculation on the origin of the  
differential-reddening corrected profile width would require a technically challenging data reduction 
to include the {\em HST} breathing effects, which is beyond the scope of the present paper.

The relatively deep ACS photometry of Palomar\,2 was used to build its radial density profile (RDP)
(Fig.~\ref{fig7}), defined as $\sigma(R) = \frac{dN}{2\pi\,RdR}$. Besides the usual power-law component 
(in this case for $R\geq1$\,pc), the RDP is quite smooth and presents a relatively flat central region, 
probably associated with some crowdedness there (see, e.g. \citealt{M15}). To determine structural 
parameters, the RDP was fitted with the classical three-parameter \citet{King62} profile, defined as 
$\sigma(R)=\sigma_0\left[\frac{1}{\sqrt{1+(R/\rC)^2}} - \frac{1}{\sqrt{1+(\rT/\rC)^2)}}\right]^2$, where 
$\sigma_0$ is the stellar-density at the center, \rC\ and \rT\ are the core and tidal radii, respectively. 
Parameters derived are $\sigma_0 = 832\pm25$ stars pc$^{-2}$, $\rC = 0.34\pm0.01\arcmin = 2.6\pm0.1$\,pc,
and $\rT = 2.96\pm0.05\arcmin = 22.5\pm0.4$\,pc. Our value of \rC\ is about twice that given by 
\citet{Harris2010}, while only about half for \rT\, probably because the relatively limited radial range of
the ACS photometry. 

We also computed its half-light radius as $R_{HL} = 0.62\pm0.02\arcmin$ (somewhat larger than the 
0.5\arcmin given by \citealt{Harris2010}), thus implying a physical value of $R_{HL}\approx4.7$\,pc. 
Incidentally, this value, together with $M_V\approx-7.8$, put Palomar\,2 right in the middle of the 
corresponding values spanned by the Milky Way GCs (see, e.g. Fig.~6 in \citealt{Belokurov2014}). To 
put these values in context, we compare in Fig.~\ref{fig7} (bottom-left panel) Palomar\,2 with the
corresponding Milky Way GCs (values taken from \citealt{Harris2010}) assumed to be in the outer halo 
(those with Galactocentric distance $R_{GC}>15$\,Kpc, e.g. \citealt{BBBO2006}) and the remaining ones. 
Finally, the range of probable \fH\ of Palomar\,2 is compared to the corresponding distribution measured 
in Milky Way GCs (outer-halo and remaining GCs) in the bottom-right panel of Fig.~\ref{fig7}. Both
comparisons confirm that Palomar\,2 is a typical outer-halo GC.

\section{Concluding remarks}
\label{CONC}

Because of the deleterious effects associated with differential reddening, CMDs of the scarcely 
studied globular cluster Palomar\,2 have been essentially useless for a more in-depth analysis
since its discovery in the 1950s. In this work, {\em Hubble Space Telescope} Advanced Camera for 
Surveys data are used to correct the photometry in F606W and F814W of each star in the field of 
view of Palomar\,2 for differential reddening with a relatively high spatial resolution. CMDs are 
extracted in different regions across the FoV of Palomar\,2, and the differences in \EBV\ among them 
are used to build a differential reddening map. Finally, correcting the observed values of F606W and 
F814W for differential reddening of each star leads to a CMD with significantly less photometric scatter 
and, consequently, well-defined stellar sequences, especially the RGB, HB and AGB.

Parameters derived from the corrected CMD show that Palomar\,2 contains a total stellar mass of 
$M=1.4\times10^5\,\ms$, is affected by the foreground reddening $\EBV=0.93$, and is located at 
$\ds=26.1\pm1.5$\,Kpc from the Sun; its absolute magnitude is $M_V\approx-7.8$, and its age is
$13.25\pm0.12$\,Gyr. The global metallicity is $\Zo\approx0.03$, which corresponds to 
$-1.91\leq\fH\leq-1.58$ (for $0.0\leq[\alpha/Fe]\leq+0.4$). Structural parameters are the core 
and half-light radii $\rC\approx2.6$\,pc and $R_{HL}\approx4.7$\,pc, respectively. These values 
are consistent with other outer halo GCs.

\section*{Acknowledgements}
Thanks to an anonymous referee for important comments and suggestions.
The authors acknowledge support from the Brazilian Institution CNPq. 



\bibliographystyle{mnras}
\bibliography{ref}

\end{document}